\documentclass[journal,twoside,web]{ieeecolor}
\usepackage{generic}
\usepackage{cite}
\usepackage{amsmath,amssymb,amsfonts}
\usepackage{color}
\newtheorem{definition}{Definition}
\newtheorem{assumption}{Assumption}
\newtheorem{theorem}{Theorem}
\newtheorem{remark}{Remark}
\newtheorem{lemma}{Lemma}
\usepackage{algorithm}
\usepackage{algorithmicx}
\usepackage[noend]{algpseudocode}
\usepackage{bm}
\usepackage{caption}
\usepackage{graphicx}
\usepackage{footmisc}
\usepackage{epstopdf}
\usepackage{textcomp}
\usepackage{balance}
\usepackage{hyperref}
\allowdisplaybreaks[4]

\def\BibTeX{{\rm B\kern-.05em{\sc i\kern-.025em b}\kern-.08em
    T\kern-.1667em\lower.7ex\hbox{E}\kern-.125emX}}
\begin{document}
\title{Communication-efficient and Differentially-private Distributed Nash Equilibrium Seeking with Linear Convergence}
\author{Xiaomeng Chen, Wei Huo, Kemi Ding, Subhrakanti Dey, and Ling Shi
\thanks{X. Chen, W. Huo and L. Shi are with the Department of Electronic and Computer Engineering, Hong Kong University of Science and Technology, Clear Water Bay, Kowloon, Hong Kong (email: xchendu@connect.ust.hk, whuoaa@connect.ust.hk, eesling@ust.hk).}
\thanks{K. Ding  is with the School of System Design and Intelligent Manufacturing, Southern University of Science and Technology, Shenzhen, 518055, China
(email: dingkm@sustech.edu.cn).} 
\thanks{S. Dey is with the Department of Electrical Engineering, Uppsala University, 751 03 Uppsala, Sweden. (email: subhrakanti.dey@angstrom.uu.se).} 
}

\maketitle
\thispagestyle{empty} 
\begin{abstract}
The distributed computation of a Nash equilibrium (NE)
for non-cooperative games is gaining increased attention recently. 	Due to the nature of distributed systems,  privacy and communication efficiency are two critical concerns. Traditional approaches often address these critical concerns in isolation.
This work introduces a unified framework, named CDP-NES, designed to improve communication efficiency in the privacy-preserving NE seeking algorithm for distributed non-cooperative games over directed graphs. Leveraging both general compression operators and the  noise adding mechanism, CDP-NES  perturbs local states with Laplacian noise and applies difference compression prior to their exchange among neighbors. We prove that CDP-NES not only achieves  {\color{black}linear convergence to a  neighborhood of the NE} in games with restricted monotone mappings but also guarantees $\epsilon$-differential privacy, addressing  privacy and communication efficiency  simultaneously.  Finally,  simulations are provided to illustrate the effectiveness of the proposed method. 
\end{abstract}

\section{INTRODUCTION}

\IEEEPARstart{N}{on-cooperative} games have been extensively studied due to their broad applications in various fields, including   congestion control in traffic networks \cite{ma2011decentralized}, charging of electric vehicles \cite{grammatico2017dynamic} and demand-side management in smart grids \cite{saad2012game}. In such games, each  agent is endowed with a local cost function to be optimized, which depends on other agents’ decisions.  The NE has emerged as a critical solution concept, reflecting the outcomes of decision-making by multiple  agents in a non-cooperative setting.  However, in 
decentralized networks where no central server exists,
individual agents do not have  access to the joint action profile to optimize its decision. As a result, agents
should share their local information with neighbors
in a distributed manner to reach the NE.

Although significant advancements have been made in distributed NE seeking \cite{koshal2016distributed,salehisadaghiani2016distributed,tatarenko2020geometric1}, these
methods have a spot in common that they all involve agents broadcasting messages. However, directly exchanging  data  will raise  two main challenges: (1) the absence of  privacy protections, risking vulnerability to inference attacks \cite{dong2015differential,gerhart2020proposition}, and (2) significant communication costs associated with transmitting high-dimensional states in large-scale systems. These issues often co-occur in distributed systems, highlighting the need for integrated solutions that address both privacy and communication efficiency.
%

Currently, communication efficiency and privacy issues are often explored separately in  distributed NE seeking. In terms of privacy, various strategies have been implemented, such as adding diminishing Laplacian noise \cite{ye2021differentially} and using globally correlated perturbation \cite{lin2023statistical} to conceal agents' actions, though these measures may compromise convergence accuracy. To deal with the tradeoff between accuracy and privacy, Wang et al. \cite{wang2022differentially}
 proposed a differentially private (DP) NE seeking method for stochastic aggregative games by perturbing the gradient used for
optimization instead of transmitting data. For general non-cooperative games, Wang and Basar \cite{wang2022ensuring} developed an approach that ensures both provable convergence and differential privacy by employing diminishing step sizes, albeit at the cost of reduced convergence speed.

In terms of communication efficiency, event-triggered mechanisms  have been suggested in the literature to
reduce the frequency of information exchanges between agents for distributed NE seeking \cite{shi2019distributed,huo2023distributed}. Furthermore, the efficiency of communication can be further improved by reducing the size of the transmitted data. This is achieved through various compression techniques, including uniform  quantizations \cite{chen2022distributed}, deterministic quantizations
\cite{ye2022distributed},  and general compressors \cite{chen2022linear}. A recent algorithm \cite{chen2023efficient} notably improved communication efficiency in distributed NE seeking problems by combining reduced transmission frequency and a reduction in data size. While effective in reducing communication overhead, these methods have not simultaneously tackled privacy concerns within their designs. {\color{black}Recently, Xie et al. \cite{XIE20238369} proposed a compressed differentially private gradient descent algorithm for distributed optimization (DO) problems on undirected graphs. However, privacy-preserving NE seeking differs fundamentally: (1) agents in DO are cooperative in  minimizing the sum of functions, whereas players in games are competitive in striving to minimize their distinct cost functions, which depends on the actions of others;  (2) noise in games can easily alter the NE (as shown in \cite{ye2021differentially}), necessitating careful design of the noise-adding mechanism and inter-player interaction to ensure convergence.}

	Motivated by the above observations, this paper designs a Compressed, Differentially Private Nash Equilibrium Seeking (CDP-NES) algorithm,  a novel approach designed to enhance communication efficiency  and preserve privacy simultaneously. {\color{black} Instead of naively compressing the transmitted data in existing privacy-preserving NE seeking methods that use diminishing stepsizes, our approach uses a difference compression technique inspired by our previous work \cite{chen2022linear} with constant stepsizes to enhance both communication efficiency and convergence speed. However, the introduction of several variables within the compression framework \cite{chen2022linear} presents challenges for integrating noise injection for privacy. Improper  noise injection  could result in algorithm divergence or privacy breaches. To address these issues, our   paper  proposes a joint design of difference compression and noise injection techniques that ensure linear convergence while preserving privacy.}  The main contributions are as follows:

 \begin{itemize}
	\item  We propose a novel algorithm (CDP-NES)  that integrates differential privacy within a compressed communication framework for distributed non-cooperative games over directed graphs.  By jointly designing  the difference compression method for noise-perturbed data, our approach can be applied to a general class of compressors, instead of the specific compressors in \cite{chen2022distributed,ye2022distributed}. 
	
	\item  By employing a constant stepsize, our approach achieves linear convergence to a neighborhood of the NE and achieves $\epsilon$-differential privacy simultaneously, significantly enhancing the convergence speed  compared to  methods that use diminishing step size \cite{ye2021differentially,lin2023statistical,wang2022differentially,wang2022ensuring}.
	
	{\color{black}	\item Moreover, we demonstrate that the mean squared error of the convergence is upper bounded by a finite value inversely proportional to the privacy level $\epsilon$, providing a clear quantification of the trade-off involved. }
\end{itemize}

 \noindent\textit{Notations:} We denote $\bm{e}_i \in\mathbb{R}^n$ as the $i$-th unit vector which takes zero except for the $i$-th entry that equals to $1$. The spectral radius of matrix $\mathbf{A}$ is denoted by $\rho(\mathbf{A})$.  Given a matrix $\mathbf{A}$, we denote its
element in the $i$-th row and $j$-th column by $[\bm{A}]_{ij}$. We denote  $||\mathbf{x}||$ and $||\mathbf{X}||_{\text{F}}$ respectively as $l_2$ norm of vector $\mathbf{x}$ and the Frobenius norm of matrix $\mathbf{X}$.  $\bm{x} \preceq \bm{y}$ is denoted as component-wise inequality between vectors $\bm{x}$ and $\bm{y}$. The smallest nonzero eigenvalue of a positive semidefinite matrix $ \mathbf{M}\neq \mathbf{0}$ is denoted as $\tilde \lambda_{\min}(\mathbf{M})$.  For a vector $\mathbf{a} \in \mathbb{R}^{n}$, we denote $\text{Diag}(\mathbf{a})$ as the diagonal matrix with the vector $\mathbf{a}$ on its diagonal. For a given constant $\theta>0$, Lap($\theta$) is the Laplace distribution with probability function $f_L(x,\theta) =\frac{1}{2\theta}e^{-\frac{|x|}{\theta}}$.  $\mathbb{E}(x)$  denotes the expectation of a random variable $x$.  

\section{PRELIMINARIES AND PROBLEM STATEMENT}
\subsection{Network Model}
We consider a group of agents  communicating  over a directed graph $G\triangleq (\mathcal{V}, \mathcal{E})$, where $\mathcal{V}=\{1,2,3\ldots, n\}$ denotes the agent set and $\mathcal{E}\subset \mathcal{V} \times \mathcal{V}$ denotes the edge set, respectively. A communication link from agent $j$ to agent $i$ is denoted by $(j,i)\in \mathcal{ E}$. The agents who can  send messages to node $i$ are denoted as in-neighbours of agent $i$ and the set of these agents is denoted as $\mathcal{N}_i^{-}=\{j\in \mathcal{V}\mid (j,i)\in \mathcal{E}\}$. Similarly,  the agents who can  receive messages from node $i$ are denoted as out-neighbours of agent $i$ and the set of these agents is denoted as $\mathcal{N}_i^{+}=\{j\in \mathcal{V}\mid (i,j)\in \mathcal{E}\}$.  The adjacency matrix of the graph is denoted as $W=[w_{ij}]_{n\times n}$ with $w_{ij}>0$ if $(i,j)\in \mathcal{ E}$ or $i=j$, and $w_{ij}=0$ otherwise. The graph $G$ is called strongly connected if there exists at least
one directed path from any agent $i$ to any agent $j$ in the graph with $i\neq j$. 
\begin{assumption}\label{asp3}
	The directed graph $G\triangleq (\mathcal{V}, \mathcal{E})$ is strongly connected. Moreover, the adjacency matrix $W$ associated with $G$ is row-stochastic, i.e., $\sum_{l=1}^n w_{il}=1, \forall i \in \mathcal{V}$.
\end{assumption}

\subsection{Compressors}
To improve the communication efficiency, we consider
the situation where agents compress the information before
sending it.  For any $\mathbf{x} \in  \mathbb{R}^d$, we  consider
a general class of stochastic compressors $\mathcal{C}(\mathbf{x}, \zeta)$  and use $f_c(\mathbf{x}, \zeta)$ to denote the corresponding probability density functions,
where ${\zeta}$ is a  random variable. Note that the realizations $\zeta$ of the compressor $\mathcal{C}(\cdot)$ are independent  among different agents and time steps. Hereafter, we drop $\zeta$ and write $\mathcal{C}(\mathbf{x})$ for notational simplicity.

\begin{assumption}\label{c3}
The compressor  $\mathcal{C}(\cdot): \mathbb{R}^d \rightarrow \mathbb{R}^d$ satisfies
	\begin{equation}\label{asp:c}
		\mathbb{E}[||\mathcal{C}(\mathbf{x} )-\mathbf{x}||^2]\leq C||\mathbf{x}||^2,\qquad \forall \mathbf{x}\in\mathbb{R}^d,
	\end{equation}
	where $C\geq 0$ and the $r$-scaling of $\mathcal{C}(\cdot)$ satisfies
	\begin{equation}\label{eq:2}
		\mathbb{E}[||\mathcal{C}(\mathbf{x})/r-\mathbf{x}||^2]\leq (1-\delta)||\mathbf{x}||^2,\qquad \forall \mathbf{x}\in\mathbb{R}^d, 
	\end{equation}
	where   $\delta \in (0,1]$ and $r>0$.
\end{assumption}
	{\color{black}	\begin{remark}
		Equation \eqref{eq:2} is  equivalent to  Equation \eqref{asp:c} with   $C = 2r^2(1-\delta) + 2(1-r)^2$ \cite{yi2023communication}. The $r$-scaling version of compressor  $\mathcal{C}(\cdot)$ ensures the  compression ratio, $1-\delta$, stays within the range of $[0,1)$ even when $C> 1$.
\end{remark}	}

The above class of compressors incorporate the widely used unbiased compressors and biased but contractive compressors, such as stochastic quantization \cite{liu2021linear} and Top-$k$  \cite{beznosikov2020biased}. They also cover non-contractive compressors like the norm-sign \cite{chen2022linear}.

\subsection{Differential Privacy}
Differential privacy serves as a mathematical notion which quantifies the degree of  involved individuals' privacy. We give the following definitions for differential privacy. 
\begin{definition}\label{df_adj}
	(Adjacency \cite{chen2023differentially})  Two function sets $\mathcal{S}^{(1)}=\{J_i^{(1)}\}^n_{i=1}$ and  $\mathcal{S}^{(2)}=\{J_i^{(2)}\}^n_{i=1}$, where {\color{black}$J_i^{(l)}: \mathbb{R}^n \rightarrow \mathbb{R}, \forall i\in\{1,2,\ldots n\},l\in{1,2}, $}  are said to be adjacent if there exists some $i_0 \in\{1,2\ldots,n\}$ such that  {\color{black}$J_i^{(1)}(x)=J_i^{(2)}(x), \forall i\neq i_0$ and $J_{i_0}^{(1)}(x)\neq J_{i_0}^{(2)}(x)$  for all $x\in \mathbb{R}^n$.}
\end{definition}


\begin{definition}\label{df_dp}
	(Differential privacy \cite{chen2023differentially}) Given $\epsilon>0,$ for any pair of adjacent function sets $\mathcal{S}^{(1)}$ and $\mathcal{S}^{(2)}$ and any observation $\mathcal{O}\subseteq \text{Range}(\mathcal{A})$, a randomized algorithm $\mathcal{A}$ keeps $\epsilon$-differential privacy if
	$P\{\mathcal{A}(\mathcal{S}^{(1)})\in \mathcal{O}\}\leq e^\epsilon P\{\mathcal{A}(\mathcal{S}^{(2)})\in\mathcal{O}\}$,
	where $\text{Range}(\mathcal{A})$ denotes the output codomain of $\mathcal{A}$.
\end{definition}

\subsection{Problem Definition}
Consider a  noncooperative game  in a multi-agent system of $n$ agents, where 
each agent has an unconstrained action set $\Omega_i=\mathbb{R}$. Without loss of generality, we assume that  each agent's decision variable is a scalar. Let $J_i$ denote  the cost function of the agent $i$. Then, the game is denoted as $\Gamma(n, \{J_{i}\}, \{\Omega_i\})$.   The goal of each agent $i\in \mathcal{V}$ is to minimize its objective function $J_i(x_i,x_{-i})$, which depends on both the local variable $x_i$ and the decision variables of the other agents $x_{-i}$. {\color{black}The concept of NE is given below. 
\begin{definition} 
	A vector $\mathbf{x}^\star=[x_1^\star,x_2^\star,\ldots,x_n^\star]^\top $ is an NE if for any $i \in \mathcal{V}$ and $x_i \in\mathbb{R}$,
$
		J_i(x_i^\star, x_{-i}^\star)\leq J_i(x_i, x_{-i}^\star).
$
\end{definition}}

 We then make the following assumptions about game $\Gamma$. 
\begin{assumption}\label{cvx}
For all $i \in\mathcal{V}$, the  cost function $J_i(x_i,x_{-i})$ is strongly convex and continuously differentiable in $x_i$ for  fixed $x_{-i}$. 
\end{assumption}

\begin{definition}
	The mapping $\mathbf{F}: \mathbb{R}^n \rightarrow \mathbb{R}^n$ of the form 
\begin{equation}
	\mathbf{F(x)}\triangleq [\nabla_1 J_1(x_1,x_{-1}),\ldots, \nabla_n J_n(x_n,x_{-n})]^\top,
\end{equation} 
is called  the game mapping of $\Gamma(n, \{J_{i}\}, \{\Omega_i \})$,
where $\nabla_i J_i(x_i,x_{-i})=\frac{\partial J_i(x_i,x_{-i})}{\partial x_i},\forall i \in\mathcal{V}$. 
\end{definition}

{\color{black}\begin{assumption}\label{mono}
	The game mapping $\mathbf{F(x)}$ is restricted strongly monotone  to any NE $\mathbf{x}^\star $ with constant $\mu_r>0$, i.e.,
	$$\langle \mathbf{F(x)}-\mathbf{F(\mathbf{x}^\star)},\mathbf{x}-\mathbf{\mathbf{x}^\star}  \rangle \geq \mu_r ||\mathbf{x-\mathbf{x}^\star}||^2, \forall \mathbf{x} \in\mathbb{R}^n.$$
		 
\end{assumption}

{\color{black}In this paper, we relax the strongly monotone mapping condition in }\cite{yu2017distributed,ye2021differentially,wang2022differentially}  to a restricted one as in Assumption \ref{mono}. Thus, a wider variety of games are taken into consideration \cite{tatarenko2020geometric1}. }

\begin{assumption}\label{asp2}
	Each function $\nabla_i J_i(x_i,x_{-i}) $ is uniformly Lipschitz continuous,  i.e., there exists a fixed constant  $ L_i \geq 0$ such that 	$\|\nabla_i J_i(\mathbf{x})-\nabla_i J_i(\mathbf{y}) \|\leq L_i \|\mathbf{x}-\mathbf{y}\|$ for all $\mathbf{x}, \mathbf{y}\in \mathbb{R}^n$.
 
\end{assumption}


In this paper, we are interested in distributed seeking of an NE in a game $\Gamma(n, \{J_{i}\}, \{\Omega_i =\mathbb{R}\})$ where Assumptions \ref{asp3},\ref{cvx},\ref{mono} and \ref{asp2} hold. Note that given  Assumptions \ref{cvx} and \ref{mono}, there exists a unique NE in the game $\Gamma(n, \{J_{i}\}, \{\Omega_i=\mathbb{R}\})$ \cite{chen2022linear}. To be specific,  $\mathbf{x}^\star \in \mathbb{R}^n$ is an NE if and only if  $\mathbf{F(x^\star)}=0. $

\section{CDP-NES Algorithm} 
In this section, we introduce CDP-NES, the proposed differentially private  NE seeking algorithm with compression  communication, where compression is achieved by   the compressors $\mathcal{C}(\cdot)$ satisfying Assumption \ref{c3}. In a distributed network, we assume that each agent maintains a local variable $
	\mathbf{x}_{(i)}=[\tilde x_{(i)1},\ldots,\tilde x_{(i)i-1},x_i,\tilde x_{(i)i+1},\ldots,\tilde x_{(i)n}]^\top \in \mathbb{R}^n,$
 which is his estimation of the joint action profile $\mathbf{x}=[x_1,x_2,
 \ldots,x_n]^\top$, where $\tilde x_{(i)j}$ denotes agent $i$'s estimate of $x_j$ and $\tilde x_{(i)i}=x_i$. 
 
 We describe  CDP-NES  in Algorithm \ref{alg1}. Different from most differentially private algorithms that directly inject noise on the transmitted data, to improve communication efficiency and achieve differential privacy, we first add noise $\xi_{i}^k$ to the local state $\mathbf{x}_{(i)}^k$ of each agent $i$, thus generating a noisy version $\mathbf{\tilde x}^k_{(i)}$. {\color{black}We then introduce  an auxiliary variable $\mathbf{h}_{i}^k$ as a reference point of $\mathbf{\tilde x}_{(i)}^k$, initialized arbitrarily in $\mathbb{R}^n$. 	Compression  is applied to the difference between   $\mathbf{\tilde x}_{(i)}^k$ and $\mathbf{h}_{i}^k$ and then we transmit the compressed data $\mathbf{q}_{i}^k$.} This ensures that only compressed information is transmitted between agents, effectively reducing the bandwidth used. {\color{black}Additionally,   as $\mathbf{h}_{i}^k$ approaches $\mathbf{\tilde x}_{(i)}^k$, the compression error correspondingly diminishes based on Assumption \ref{c3}. The introduction of  $\mathbf{h}_{i}^k$ is motivated from  DIANA \cite{mishchenko2019distributed}, which  eliminates the compression error, particularly for a relatively large  constant $C$ in Assumption  \ref{c3}.}

  After receiving the compressed value, each agent $i$ constructs  an estimation of $\mathbf{x}_{(i)}$, denoted as $\mathbf{\hat x}_{(i)}$, and utilizes this estimation to update their local variable 
 $\mathbf{x}_{(i)}^k$  for mitigating the compression error.  This innovative mechanism is designed to ensure linear convergence and maintain differential privacy within a compressed communication framework, with its effectiveness theoretically validated in Sections \ref{con}--\ref{pr}. 


%

 \begin{algorithm}[t]
	\caption{A Compressed and Differentially Private Distributed Nash Equilibrium Seeking (CDP-NES) Algorithm}
	\label{alg1}
 {\bf Input:} Stopping time $K$, stepsize $\eta$, consensus stepsize $\gamma$, scaling parameter $\alpha>0$, noise variance $\theta_i>0$ and initial values $\mathbf{x}_{(i)}^0, \mathbf{h}_{i}^0$  \\
	 {\bf Output: $\mathbf{x}_{(i)}^K$}
	\begin{algorithmic}[1]
		\For {each agent $i\in\mathcal{V}$}
		\State $\mathbf{h}_{i,w}^0=\sum\limits_{j\in \mathcal{N}_i} w_{ij}\mathbf{h}_{j}^0$
		\EndFor
		\For {$k=0,1,2,\ldots, K-1$} locally at each agent $i\in\mathcal{V}$ 
		\State $\mathbf{\tilde x}^k_{(i)} = \mathbf{x}_{(i)}^k+\xi_{i}^k$, where 
	$\xi_{i}^k$ is a random vector consisting of $n$  Laplacian noise independently drawn from Lap($\theta_i$)  \hfill \Comment{Noise perturbation}
		\State  
		$\mathbf{q}_i^k=\mathcal{C}(\mathbf{\tilde x}_{(i)}^k-\mathbf{h}_{i}^k)$  \Comment{Difference compression}
		\State $\mathbf{\hat x}_{(i)}^k=\mathbf{h}_{i}^k+\mathbf{q}_i^k$
		\State $\mathbf{h}_{i}^{k+1}=(1-\alpha)\mathbf{h}_{i}^k+\alpha \mathbf{\hat x}_{(i)}^k$
		\State Send $\mathbf{q}_i^k$ to agent $l \in \mathcal{N}_i^-$ and then receive $\mathbf{q}_j^k$ from agent $j \in \mathcal{N}_i^+$ \Comment{Communication}
		\State $\mathbf{\hat x}_{(i),w}^k =\mathbf{h}_{i,w}^k+\sum_{j=1}^n w_{ij}\mathbf{q}_j^k$
		\State $\mathbf{h}_{i,w}^{k+1}=(1-\alpha)\mathbf{h}_{i,w}^k+\alpha\mathbf{\hat x}_{(i),w}^k$
		\State $\mathbf{x}_{(i)}^{k+1}=\mathbf{\tilde x}_{(i)}^{k}-\gamma(\mathbf{\hat x}_{(i)}^k-\mathbf{\hat x}_{(i),w}^k)-\gamma\eta\nabla_i J_i(\mathbf{x}_{(i)}^{k})\mathbf{e}_i$
	\EndFor
\end{algorithmic}
\end{algorithm}


Denote the compact form of local state variable  as $\mathbf{X}=[\mathbf{x}_{(1)},\mathbf{x}_{(2 )},\ldots,\mathbf{x}_{(n)}]^\top\in \mathbb{R}^{n \times n},$
where the $i$th row is  the  vector $\mathbf{x}_{(i)}, i\in\mathcal{V}$. Auxiliary variables of the agents in compact form $\bm{\xi}, \mathbf{\tilde X}, \mathbf{H},\mathbf{H}_w, \mathbf{Q}, \mathbf{\hat X}$ and $ \mathbf{\hat X}_w$ are defined similarly.  At $k-$th iteration, their values are denoted by $\mathbf{X}^k,\bm{\xi}^k, \mathbf{\tilde X}^k, \mathbf{H}^k,\mathbf{H}_{w}^k, \mathbf{Q}^k, \mathbf{\hat X}^k$ and $ \mathbf{\hat X}_{w}^k$ , respectively. 
Moreover, for any given action-profile estimates, we define a diagonal matrix 
$\mathbf{\tilde F}(\mathbf{X})\triangleq \text{Diag}(\nabla_1 J_1(\mathbf{x}_{(1)}),\ldots, \nabla_n J_n(\mathbf{x}_{(n)})).$

 Algorithm \ref{alg1} can be written in compact form as follows: 
\vspace*{-2mm}
\begin{subequations}
	\begin{align}\label{ala}
		&\mathbf{\tilde X}^{k}=\mathbf{X}^{k}+\bm{\xi}^k,\\ 
		&\mathbf{Q}^k=\mathcal{C}(\mathbf{\tilde X}^{k}-\mathbf{H}^{k}),\\ \label{alb} 
		&\mathbf{\hat X}^{k}=\mathbf{H}^{k}+\mathbf{Q}^k,\\ 
		&\mathbf{\hat X}^{k}_w=\mathbf{H}_w^{k}+W\mathbf{Q}^k,\\\label{ald}
		&\mathbf{H}^{k+1}=(1-\alpha)\mathbf{H}^{k}+\alpha\mathbf{\hat X}^{k},\\
		&\mathbf{H}^{k+1}_w=(1-\alpha)\mathbf{H}^{k}_w+\alpha\mathbf{\hat X}^{k}_w,\\
		&\mathbf{X}^{k+1}=\mathbf{\tilde X}^{k}-\gamma(\mathbf{\hat X}^{k}-\mathbf{\hat X_w}^{k})-\gamma\eta \mathbf{\tilde F}(\mathbf{X}^{k}),\label{alf} 
	\end{align}
\end{subequations}
where $\mathbf{X}^{0}$ and $\mathbf{H}^{0}$ are arbitrary chosen.

Since $\mathbf{H}^{k}_w=W\mathbf{H}^{k}$ and $\mathbf{\hat X}^{k}_w=W\mathbf{\hat X}^{k}$  for all $k$ \cite{chen2022linear},   the state variable update in \eqref{alf} becomes
\begin{equation}\label{algx}
	\begin{aligned}
		&\mathbf{X}^{k+1}=\mathbf{\tilde X}^{k}-\gamma(\mathbf{\hat X}^{k}-W\mathbf{\hat X}^{k})-\gamma\eta \mathbf{\tilde F}(\mathbf{X}^{k}) \\
		&=\mathbf{ X}^{k}-\gamma \mathbf{F}_a(\mathbf{X}^{k})+\gamma(I-W)\mathbf{ E}^{k}+\bm{\xi}^k,\\
	\end{aligned}
\end{equation}where $\mathbf{F}_a(\mathbf{X}^k)=(I-W)\mathbf{X}^k+\eta\mathbf{\tilde F}(\mathbf{X}^k)$ denotes  the \textit{augmented mapping}  \cite{tatarenko2020geometric1} of game $\Gamma$, and $\mathbf{ E}^{k}=\mathbf{X}^{k}-\mathbf{\hat X}^{k}$.

\section{Convergence Analysis}\label{con}
In this section, we analyze the convergence performance of CDP-NES. The main idea of our strategy is to bound the NE-seeking error 	$\mathbb{E}[||\mathbf{X}^{k}-\mathbf{X}^\star||^2_{\text{F}}]$ and the compression error $\mathbb{E}[||\mathbf{\tilde X}^{k}-\mathbf{H}^{k}||^2_{\text{F}}]$ on the
basis of the linear combinations of their previous values. Let $\mathcal{F}^k$ be the $\sigma-$algebra generated by $\{\mathbf{E}^{0},\bm{\xi}^{0}, \mathbf{ E}^{1}, \bm{\xi}^{1},\ldots, \mathbf{E}^{k-1},\bm{\xi}^{k-1}\}$, and denote $\mathbb{E}[\cdot| \mathcal{F}^k]$ as the conditional expectation  given $\mathcal{F}^k$. 

\subsection{Supporting Lemmas}

\begin{lemma} \label{lml}(Lemma 1 in \cite{tatarenko2020geometric1})
	Given Assumption \ref{asp2}, the augmented mapping $\mathbf{F}_a$	of game $\Gamma$ is Lipschitz continuous  with  $L_F=\eta L_m+||I-W||_{\text{F}}$, where $L_m=\max_iL_i$. 
\end{lemma}
\begin{lemma}\label{lmmu} (Lemma 3 in \cite{tatarenko2020geometric1})
	Given Assumptions \ref{asp3}, \ref{mono} and \ref{asp2}, the augmented mapping $\mathbf{F}_a$	of game $\Gamma(n, \{J_{i}\}, \{\Omega_i\})$ is restricted strongly monotone to any NE  $\mathbf{X}^\star$ with the constant $\mu_F=\min\{b_1,b_2\}>0$, where $b_1=\eta\mu_r/2n, b_2=(\beta^2\tilde \lambda_{\text{min}}(I-W)/(\beta^2+1))-\eta^2 L_m$ and $\beta$ is a positive constant such that $\beta^2+2\beta=\frac{\mu_r}{2n\eta L_m}$.  
\end{lemma}

\subsection{Main results}


\begin{lemma}\label{lmm}
	Suppose Assumption \ref{c3} holds, the variables $\mathbf{X}^k, \mathbf{H}^k$ are measurable  respect to $\mathcal{F}^k$. Furthermore, we have 
	\begin{equation}
		\mathbb{E}[||\mathbf{E}^k||^2_{\text{F}}\mid \mathcal{F}^{k}]\leq  2C||\mathbf{\tilde X}^k-\mathbf{H}^k||^2_{\text{F}}+4n^2\bar \theta^2,
	\end{equation}
	where $\bar \theta=\max_{i \in \mathcal{V}}\theta_i$ and $\theta_i$ denotes the Laplacian noise variance of agent $i$.
\end{lemma}

\textit{Proof:}	By expanding \eqref{ala}-\eqref{alf} recursively, $\mathbf{X}^k$ and $ \mathbf{H}^k$ can be  represented by linear combinations of $\mathbf{ X}^{0}, \mathbf{H}^{0}$ and random variables $\{\mathbf{E}^j\}_{j=0}^{k-1}$ and $\{\bm{\xi}^j\}_{j=0}^{k-1}$, i.e, $\mathbf{X}^k, \mathbf{H}^k$ are measurable  respect to $\mathcal{F}^k$. Moreover, from  \eqref{alb} and \eqref{asp3}, we have  
	$$	\begin{aligned}
			&\mathbb{E}[||\mathbf{E}^k||^2_{\text{F}}\mid \mathcal{F}^{k}]=\mathbb{E}[||\mathbf{X}^k-\mathbf{\hat X}^k||^2_{\text{F}}\mid \mathcal{F}^{k}]\\
			&=\mathbb{E}[||\mathbf{\tilde X}^k-\bm{\xi}^k-\mathbf{H}^k-\mathcal{C}(\mathbf{\tilde X}^k-\mathbf{H}^k)||^2_{\text{F}}\mid \mathcal{F}^{k}]\\
			&\leq 2 \mathbb{E}[||\mathbf{\tilde X}^k-\mathbf{H}^k-\mathcal{C}(\mathbf{\tilde X}^k-\mathbf{H}^k)||^2_{\text{F}}\mid\mathcal{F}^{k}]+2\mathbb{E}[||\bm{\xi}^k||^2_{\text{F}}\mid\mathcal{F}^{k}]\\
			&\leq 2C||\mathbf{\tilde X}^k-\mathbf{H}^k||^2_{\text{F}}+4n^2\bar \theta^2.\qquad\qquad\qquad\qquad\qquad\qquad{\hfill \blacksquare}		\end{aligned} 	$$

	 \begin{lemma}\label{mainl}
		Given Assumptions \ref{asp3}--\ref{asp2}, when $\alpha \in(0,\frac{1}{r}]$, the following linear system of  component-wise inequalities holds   
		\begin{equation}\label{inequlity}
			\mathbf{V}^{k+1}\preceq \mathbf{A}\mathbf{V}^{k}+\mathbf{b},\
		\end{equation}	
			where 
	$
			\mathbf{V}^{k}=
			\begin{bmatrix}
				\mathbb{E}[||\mathbf{X}^{k}-\mathbf{X}^\star||^2_{\text{F}}\mid \mathcal{F}^k]&\mathbb{E}[||\mathbf{\tilde X}^{k}-\mathbf{H}^{k}||^2_{\text{F}}\mid \mathcal{F}^k]\\
			\end{bmatrix}^\top	
$
		and the  elements of  $\mathbf{A}$ and  $\mathbf{b} $ are respectively given by
		
		\begin{equation}
			\begin{aligned}
				&\mathbf{A}=\begin{bmatrix}
					c_1(1+L_F^2\gamma^2-2\mu_F\gamma)&c_2\gamma^2\\
					c_4\gamma^2 L_F^2&c_x+c_5\gamma^2\\
				\end{bmatrix},
				\\
				&\mathbf{b}=\begin{bmatrix}
					c_3\bar\theta^2&c_6\bar\theta^2
				\end{bmatrix}^\top,	
			\end{aligned}
		\end{equation}
		where $c_1=\frac{2L_F^2-\mu_F^2}{2L_F^2-2\mu_F^2},c_2=\frac{4c_1||I-W||_{\text{F}}^2C}{c_1-1}, c_3=\frac{8c_1||I-W||_{\text{F}}^2n^2}{c_1-1}+\frac{4c_1n^2}{c_1-1}, c_4=\frac{6(1+\alpha r \delta)}{\alpha r \delta},  c_5=2c_4C||I-W†||^2_{\text{F}},  c_6=(2c_4+6)n^2$ and $c_x=1-\alpha^2r^2\delta^2$ .
		
	\end{lemma}

	\textit{Proof:} See Appendix \ref{ap1}.


	\begin{theorem}\label{th1}
		Suppose Assumptions \ref{asp3}--\ref{asp2} hold. When $\gamma=\mu_F/L_F^2$,  the  parameter $\alpha$ and gradient stepsize $\eta$ satisfy
		
		\begin{equation*}\label{pa2}
			\alpha \in\Big(0,\frac{1}{r}\Big], \quad	\eta\le \min\Big\{\frac{2n}{\mu_r}\sqrt{\frac{(1-c_x)}{m_1}},\frac{\tilde \lambda_{\text{min}}(I-W)m_2^2}{L_m(m_2^2+1)}\Big\},
		\end{equation*}
		with $m_1,m_2$ are defined in \eqref{var_m1} and \eqref{var_m2}, then $\sup_{l\geq k}\mathbb{E}[||\mathbf{ X}^{k+1}-\mathbf{X}^\star||^2_{\text{F}}]$ and $\sup_{l\geq k}\mathbb{E}[||\mathbf{\tilde X}^{k+1}-\mathbf{H}^{k+1}||^2_{\text{F}}]$ converge to $\lim\sup_{k\rightarrow\infty}\mathbb{E}[||\mathbf{ X}^{k+1}-\mathbf{X}^\star||^2_{\text{F}}]$ and  $\lim\sup_{k\rightarrow\infty} \mathbb{E}[||\mathbf{\tilde X}^{k+1}-\mathbf{H}^{k+1}||^2_{\text{F}}]$, respectively, at the linear rate $\mathcal{O}(\rho(\mathbf{A})^k)$ with $\rho(\mathbf{A})<1$. In addition, the NE-seeking error is bounded by
		\begin{equation}\label{bound_ne}
		\limsup_{k\rightarrow\infty} \mathbb{E}[||\mathbf{ X}^{k+1}-\mathbf{X}^\star||^2_{\text{F}}]\leq c_7 \bar \theta^2,	\vspace*{-2mm}
		\end{equation}
		where {\color{black}$c_7=((1-c_x-c_5\gamma^2)c_3+c_2\gamma^2)/\text{det}\mathbf{(I-A)}.$} 
		 \end{theorem} 
	
\textit{Proof:}
In terms of Lemma \ref{mainl}, by induction we have
		\begin{equation}\label{eqi}
		\mathbf{V}^{k}\preceq \mathbf{A}^k 	\mathbf{V}^{0}+\sum_{l=0}^{k-1}\mathbf{A}^l\mathbf{b}.
	\end{equation}
		From equation \eqref{eqi}, we can see that if  $\rho(\mathbf{A})<1$, then $\limsup_{k\rightarrow\infty} \mathbb{E}[||\mathbf{ X}^{k+1}-\mathbf{X}^\star||^2_{\text{F}}]$  and $\limsup_{k\rightarrow\infty} \mathbb{E}[||\mathbf{\tilde X}^{k+1}-\mathbf{H}^{k+1}||^2_{\text{F}}]$ all converge to a neighborhood of $0$ at the linear rate $\mathcal{O}(\rho(\mathbf{A})^k)$. Similar to the proof of Theorem 1 in \cite{chen2022linear}, we can obtain that 
	 $\rho(\mathbf{A})\leq(1-\frac{\mu_F^2}{4L_F^2})<1$, if the positive constants $\epsilon_1, \epsilon_2$ and the stepsize $\eta$ satisfy the following conditions, 
	\begin{equation*}
		\begin{split}
			&\epsilon_1=\frac{4c_2\epsilon_2}{L_F^2},\eta\le \min\Big\{\frac{2n}{\mu_r}\sqrt{\frac{(1-c_x)\epsilon_2}{m_1}},\frac{\tilde \lambda_{\text{min}}(I-W)m_2^2}{L_m(m_2^2+1)}\Big\},
		\end{split}
	\end{equation*}
	where
	\begin{subequations}
		\begin{align}\label{var_m1}
			&m_1=\frac{4c_2c_4}{||I-W||_{\text{F}}^4}+\frac{1}{4||I-W||_{\text{F}}^2}+\frac{c_5}{||I-W||_{\text{F}}^4},\\\label{var_m2}
			&m_2=-1+\sqrt{1+\frac{\mu_r^2}{4n^2L_m}\sqrt{\frac{m_1}{(1-c_x)\epsilon_2}}}.
		\end{align}
		\end{subequations}
	Moreover, since $\lim\limits_{k \rightarrow \infty}\sum_{l=0}^{k-1}\mathbf{A}^l=\mathbf{(I-A)}^{-1}$, we have $\limsup_{k\rightarrow\infty} \mathbb{E}[||\mathbf{ X}^{k+1}-\mathbf{X}^\star||^2_{\text{F}}]\leq [\mathbf{(I-A)^{-1}b}]_1=c_7\bar \theta^2. \hfill \blacksquare$
	

	\begin{remark}
			{\color{black}Theorem \ref{th1} shows that  CDP-NES achieves linear convergence to a neighborhood of the NE with a linear convergence rate $\mathcal{O}(\rho(\mathbf{A})^k)$, where $\rho(\mathbf{A})\leq 1-\mu_F^2/4L_F^2$.  Based on the definition of $\mu_F$ and $L_F$ in Lemmas \ref{lml} and \ref{lmmu}, the convergence rate is dependent on the stepsize $\eta$, the number of players, properties of the game, and the connectivity of the communication graph. Moreover, the convergence accuracy is inversely related to the maximum noise variance $\bar \theta$.  Specifically, a smaller $\bar \theta$ leads to a more accurate convergence point.} If there is no Laplacian noise, i.e., $\bar \theta=0$,  Algorithm \ref{alg1} converges linearly to the NE. However, as discussed in Section~\ref{pr}, reducing the noise variance compromises the degree  of differential privacy, indicating an inherent trade-off between convergence accuracy and differential privacy.
	\end{remark}

	\section{Privacy analysis}\label{pr}
In this section, we demonstrate that the CDP-NES algorithm ensures the differential privacy of all agents' cost functions.
	
	We use $\mathcal{O}^k$ to denote  the information transmitted between
	agents at time step $k$, i.e., $\mathcal{O}^k=\{\mathcal{C}(\mathbf{\tilde x}_{(i)}^k-\mathbf{h}_{i}^k)\mid \forall i \in \mathcal{V}\}$. Consider
	any two adjacent function sets $\mathcal{S}^{(1)}$ and $\mathcal{S}^{(2)}$, and only the cost function $J_{i_0}$ is different between the two sets, i.e., $J_{i_0}^{(1)}\neq J_{i_0}^{(2)}$ and $J_{i}^{(1)}=J_{i}^{(2)}, \forall i \neq i_0.$ Prior to presenting the privacy result, we introduce the following assumption:	\begin{assumption}\label{bounded_grad}
	The gradient of agent $i$’s objective function
is uniformly bounded, i.e., there exists a positive constant $M$ such that 
		$||\nabla_i J_i(\mathbf{x})||_1 \leq M$  for all $\mathbf{x} \in \mathbb{R}^n, i \in \mathcal{V}$.
	\end{assumption}
	
Assumption \ref{bounded_grad} is widely used in existing differentially private works \cite{ye2021differentially, wang2022differentially} and it is crucial for privacy analysis.

	\begin{theorem}\label{th2}
		For any finite number of iterations $K$, under Assumption \ref{bounded_grad},  CDP-NES preserves $\epsilon_i$-differential privacy for a given $\epsilon_i>0$ of each agent $i$'s cost function if the noise parameter satisfies $\theta_i >2\gamma \eta K M/\epsilon_i$, {\color{black}where $\epsilon_i$ denotes the cumulative privacy budget over $K$ iterations for agent $i$.}
		
	\end{theorem}
	
	\textit{Proof:} See Appendix \ref{ap2}. 

{\color{black}\begin{remark}
	The privacy analysis in Theorem \ref{th2} relaxes the assumption of uniform second-order gradients  (Assumption 4  in \cite{XIE20238369}), demonstrating that our privacy-preserving algorithm can be applied to a wider class of problems. Furthermore, unlike the decaying noise approach in \cite{XIE20238369}, which risks privacy as convergence is reached,  Algorithm \ref{alg1} maintains constant noise variance, enhancing privacy protection.
\end{remark}%
}
\begin{remark}
 In this paper, we consider that each agent has an unconstrained action set. If $\Omega=\Omega_1\times\Omega_2\cdots\times \Omega_n$ is a nonempty, convex and compact set, agent can update their local variable using a projection operator
  \begin{equation}\label{constraints}
  \mathbf{x}_{(i)}^{k+1}=P_{\Omega}\{\mathbf{\tilde x}_{(i)}^{k}-\gamma(\mathbf{\hat x}_{(i)}^k-\mathbf{\hat x}_{(i),w}^k)-\gamma\eta\nabla_i J_i(\mathbf{x}_{(i)}^{k})\mathbf{e}_i\}
  	  \end{equation}
  	  where  $P_{\Omega}$ denotes the Euclidean projection operator onto $\Omega$ and  updates for other variables are the same in Algorithm \ref{alg1}. In this case, the corresponding analyses are similar to the presented ones by utilizing the non-expansive property of the  projection operators \cite{tatarenko2020geometric1}. {\color{black}By imposing a constraint on $\Omega$, the bounded gradient requirement in Assumption \ref{bounded_grad} can be satisfied.} Additionally, the proposed algorithm can be directly adapted to scenarios where  agents’ actions are multi-dimensional.\end{remark}


\section{SIMULATIONS} \begin{figure}[t]  	\centering
  	\includegraphics[width=0.49\textwidth]{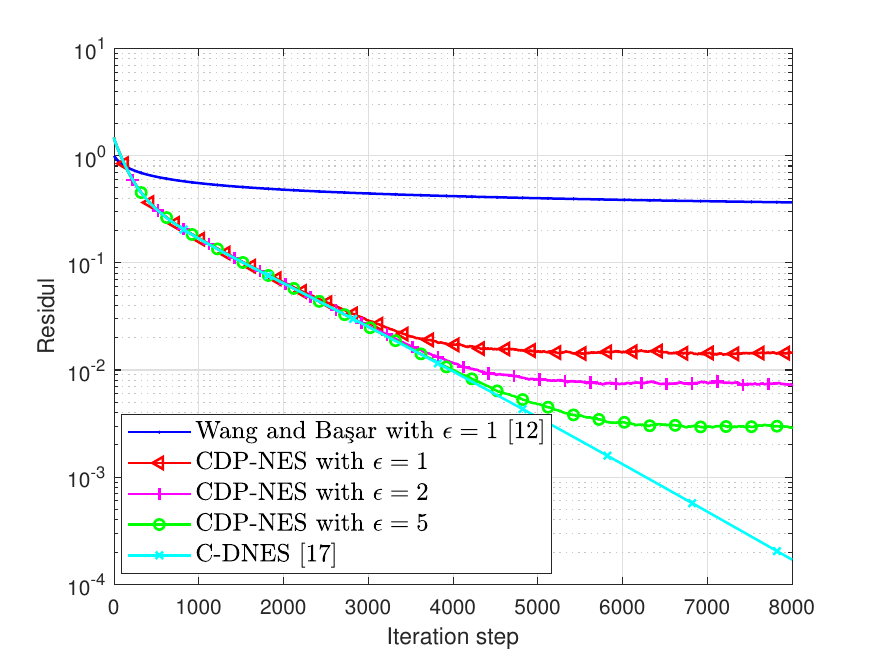}
  	\caption{Residual ($\mathcal{R}=||\mathbf{X}^k-\mathbf{X}^\star||_{\text{F}}$) v.s. number of iterations for the differential-private  NE seeking  in Wang and Ba\c{s}ar \cite{wang2022ensuring}, the compressed NE seeking (C-DNES) \cite{chen2022linear} and the proposed CDP-NES algorithm with different privacy budgets.}
  	\label{simu}
  \end{figure}

Consider a randomly generated directed communication network consisting of $n=50$ agents. The connectivity control game \cite{ye2020distributed} is considered, where the sensors  try to find a tradeoff between the local objective, e.g., source seeking, and the global objective, e.g., maintaining connectivity with other sensors. The cost function of agent $i$ is defined as follows,
\begin{equation}
\begin{aligned}
		J_i(\mathbf{x})=l_i^c(x_i)+l_i^g(\mathbf{x}),\\
\end{aligned}
	\end{equation}
	where $l_i^c(x_i)=r_{ii}x_i^\top x_i+x_i^\top r_i+b_i$, $l_i^g(\mathbf{x})=\sum\limits_{j\in S_i} c_{ij}||x_i-x_j||^2$,  {\color{black}$S_i\subseteq V$ denotes the physical neighbor set of sensor $i$}, and $x_i=[x_{i1},x_{i2}]^\top\in\mathbb{R}^2$ denotes the position of sensor $i$, $r_{ii},r_i, b_i, c_{ij}>0$ are constants. 

In this simulation, the parameters are set as $r_{ii}=b_i=i, r_i=\begin{bmatrix}i&i \end{bmatrix}^\top$ and $c_{ij}=1,    \forall i,j \in V$. Furthermore, there is $S_i=\{i+1\}$ for $i\in\{1,2,\ldots,49\}$ and $S_{50}=\{1\}$. Moreover, we consider that $x_i\in[-10,10]^2, \forall i\in\mathcal{V}$. To address the box constraints, a projection operator is included in the proposed approach as in \eqref{constraints}.
The  unique Nash equilibrium of the game is $x^\star_{ij}=-0.5$ for $i\in\{1,2,\ldots,50\}, j\in\{1,2\}$. Meanwhile,  $\mathbf{x}_{(i)}^0$ are randomly generated in $[0,1]^{100}$, $\mathbf{h}_{i}^0=\mathbf{0}$, the scaling paramater $\alpha$ and the consensus stepsize $\gamma$ are set to $0.01$. For data  compression, we adopt  the stochastic quantization compressor  with $b=2$ and $q=\infty$ as  shown in  \cite{liu2021linear}. As shown in \cite{chen2022linear}, transmitting this compressed message needs $(b+1)d+l$ bits if a scalar can be transmitted with $l$ bits while maintaining sufficient precision. In contrast, each agent sends $dl$ bits at each   iteration when adopting  NE seeking algorithm without compression. In this simulation, we choose $l=32$.

  By setting the gradient stepsize as $\eta=0.01$ and experimenting with {\color{black}  different cumulative privacy budgets $\epsilon_i=\epsilon=1,2,5, \forall i\in \mathcal{V}$ over $K=8000$ iterations}, {\color{black}Fig. \ref{simu} demonstrates that compared to the compressed NE seeking (C-DNES) \cite{chen2022linear} without privacy preservation, CDP-DNES maintains a similar linear convergence rate as C-DNES but  converges  to a neighborhood of the NE due to the effect of perturbed  noise. }Notably, a reduction in $\epsilon$ results in an increased residual error, indicating that enhanced privacy levels compromise accuracy. 

Additionally,  we compare the proposed CDP-DNES with the existing differentially-private NE seeking algorithm by Wang and Ba\c{s}ar \cite{wang2022ensuring}. By aligning the stepsize and diminishing sequence with those utilized in \cite{wang2022ensuring} and  setting the differential privacy degree  to $\epsilon=1$,  Fig. \ref{simu} illustrates that within the same time iteration, CDP-DNES  converges faster and generates smaller residual than Wang and Ba\c{s}ar \cite{wang2022ensuring}.
   
\begin{figure}[t]
	\centering
	\includegraphics[width=0.47\textwidth]{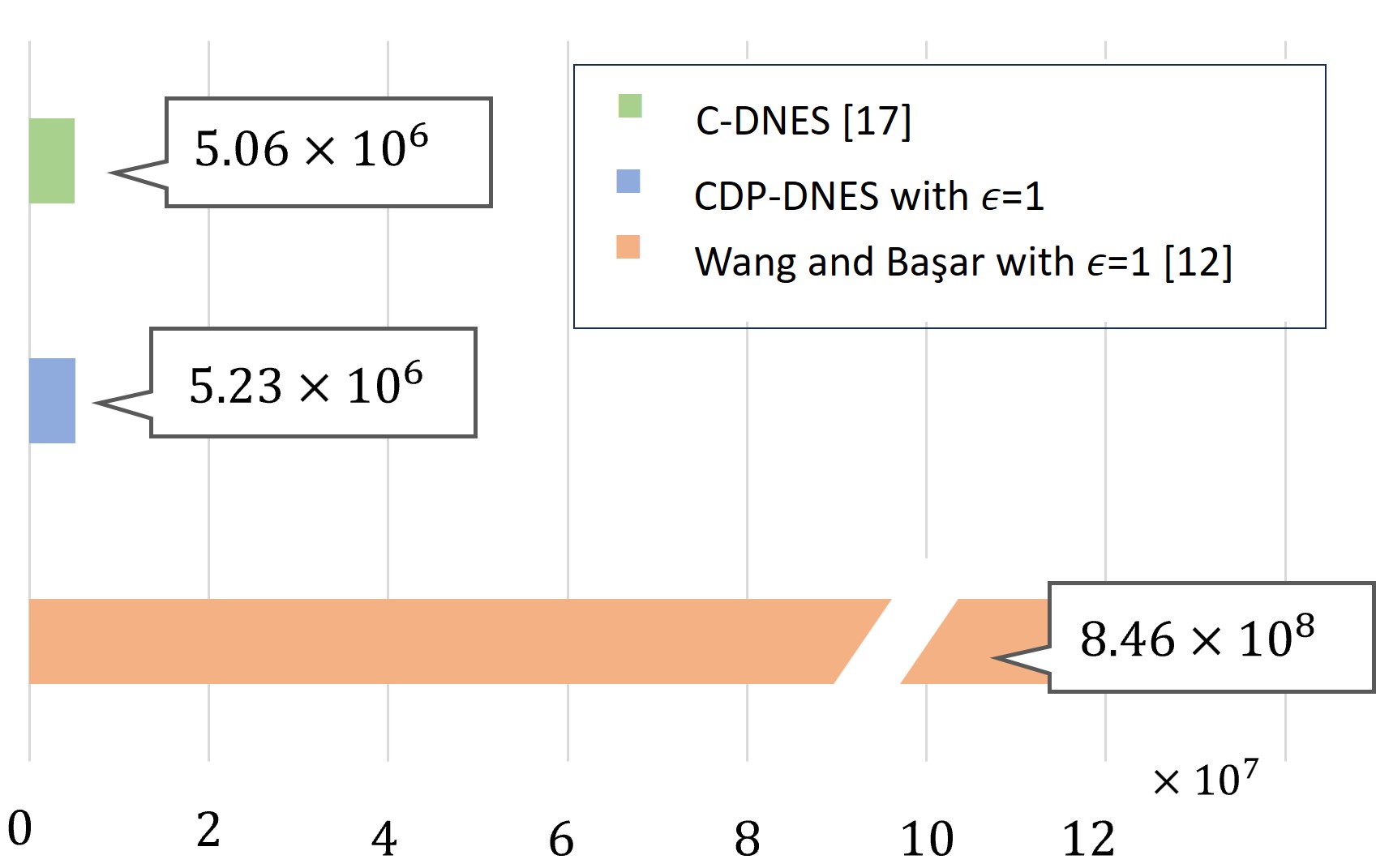}
	\caption{Total numbers of transmitted bits of different algorithms to obtain $\mathcal{R}\leq 0.02$.}
	\label{tot_comp}
\end{figure}

 {\color{black}  Furthermore, the effectiveness of  compression mechanism is illustrated in Fig. \ref{tot_comp}. To reach a specific level of error,   our proposed CDP-NES matches C-DNES in communication efficiency measured by   total number of transmitted bits. Comparing to Wang and Ba\c{s}ar's approach \cite{wang2022ensuring},   CDP-NES significantly reduces the number of transmitted bits, enhancing communication efficiency by two order of magnitude.}

\section {CONCLUSION AND FUTURE WORK}
In this paper, we   studied  differentially private distributed Nash equilibrium seeking over directed networks with compressed communication.	By injecting random noise before difference compression, a novel compressed and differentially private distributed NE seeking approach is proposed to save communication efficiency and achieve differential privacy simultaneously. Moreover, linear convergence and $\epsilon$-differential privacy is rigorously proved. Future works may focus on distributed NE seeking with coupling equality constraints.  In addition, it is of interest to rigorously investigate   the relationship between compressors and privacy performance.

\bibliographystyle{IEEEtran}
\bibliography{ref.bib}

\section{Appendix}\label{ap}
\subsection{Proof of Lemma \ref{mainl}}\label{ap1}
We derive two inequalities in terms of NE-seeking error and compression error, respectively. 

\textit{NE-seeking error:}    

Based on Proposition 1 in \cite{tatarenko2020geometric1}, we conclude that $\mathbf{X}^\star=\mathbf{X}^\star-\gamma\mathbf{F}_a(\mathbf{X}^\star), \forall \gamma>0$.  Thus,   we can obtain 
\begin{equation}\label{eqxstar1}
	\begin{aligned}
		&||\mathbf{X}^{k+1}-\mathbf{X}^\star||_{\text{F}}^2\\
		&=||\mathbf{X}^{k}-\gamma \mathbf{F}_a(\mathbf{X}^{k})+\gamma(I-W)\mathbf{ E}^{k}+\bm{\xi}^k-\mathbf{X}^\star+\gamma\mathbf{F}_a(\mathbf{X}^\star)||_{\text{F}}^2\\
		&\le c_1||\mathbf{X}^{k}-\gamma \mathbf{F}_a(\mathbf{X}^{k})-\mathbf{X}^\star+\gamma\mathbf{F}_a(\mathbf{X}^\star)||_{\text{F}}^2 \\
		&+\frac{c_1}{c_1-1}(2\gamma^2||I-W||_{\text{F}}^2||\mathbf{E}^{k}||_{\text{F}}^2+2||\bm{\xi}||^2_{\text{F}}),
	\end{aligned}
\end{equation} 
where  the last equality comes from Lemma 5 in \cite{liao2022compressed} with $c_1>1$. 

Next, we bound
\begin{equation}\label{eqxstar2}
	\begin{aligned}
		&||\mathbf{X}^{k}-\gamma \mathbf{F}_a(\mathbf{X}^{k})-\mathbf{X}^\star+\gamma\mathbf{F}_a(\mathbf{X}^\star)||_{\text{F}}^2\\
		&=||\mathbf{X}^{k}-\mathbf{X}^\star||_{\text{F}}^2+\gamma^2|| \mathbf{F}_a(\mathbf{X}^{k})-\mathbf{F}_a(\mathbf{X}^\star)||_{\text{F}}^2\\
		&-2\gamma\langle \mathbf{F}_a(\mathbf{X}^{k})-\mathbf{F}_a(\mathbf{X}^\star),\mathbf{X}^{k}-\mathbf{X}^\star\rangle\\
		&\le (1+\gamma^2L_F^2-2\gamma\mu_F)||\mathbf{X}^{k}-\mathbf{X}^\star||_{\text{F}}^2,
	\end{aligned}
\end{equation}
where the last inequality is based on Lemma \ref{lml} and Lemma \ref{lmmu}. 

From \eqref{eqxstar2},  we have $\min\{1+\gamma^2L_F^2-2\gamma\mu_F\}=(1-\mu_F^2/L_F^2)>0$. Combining \eqref{eqxstar1} and \eqref{eqxstar2} after taking taking $c_1=\frac{2L_F^2-\mu_F^2}{2L_F^2-2\mu_F^2}$, we have

\begin{equation}\label{eqxstar3}
	\begin{aligned}
		||\mathbf{X}^{k+1}-\mathbf{X}^\star||_{\text{F}}^2&\le c_1(1+\gamma^2L_F^2-2\gamma\mu_F)||\mathbf{X}^{k}-\mathbf{X}^\star||_{\text{F}}^2 \\
		&+\frac{2c_1\gamma^2||I-W||_{\text{F}}^2}{c_1-1}||\mathbf{ E}^{k}||_{\text{F}}^2+\frac{2c_1}{c_1-1}||\bm{\xi}^{k}||_{\text{F}}^2,
	\end{aligned}
\end{equation}

Then, from Lemma \ref{lmm}, we can obtain
\begin{equation}\label{eqxstar4}
	\begin{aligned}
		&\mathbb{E}[||\mathbf{X}^{k+1}-\mathbf{X}^\star||_{\text{F}}^2\mid \mathcal{F}^k]\\
		&\leq c_1(1+\gamma^2L_F^2-2\gamma\mu_F)\mathbb{E}[||\mathbf{X}^{k}-\mathbf{X}^\star||_{\text{F}}^2\mid \mathcal{F}^k] \\
		&+c_2\gamma^2\mathbb{E}[||\mathbf{\tilde X}^{k}- \mathbf{H}^{k}||^2_{\text{F}}\mid \mathcal{F}^k]+c_3\bar \theta^2,
	\end{aligned}
\end{equation}
where $c_2=\frac{4c_1||I-W||_{\text{F}}^2C}{c_1-1}, c_3=\frac{8c_1||I-W||_{\text{F}}^2n^2}{c_1-1}+\frac{4c_1n^2}{c_1-1}$. 

\textit{Compression error:} 

Denote $\mathcal{C}_r(\mathbf{X}^{k})=\mathcal{C}(\mathbf{X}^{k})/r$,  according to \eqref{ald}, for $0<\alpha\leq \frac{1}{r}$,
\begin{equation}\label{cpe1}
	\begin{split}
		&    	||\mathbf{\tilde X}^{k+1}- \mathbf{H}^{k+1}||^2_{\text{F}}\\
		=&||\mathbf{\tilde X}^{k+1}-\mathbf{\tilde X}^{k}+\mathbf{\tilde X}^{k}- \mathbf{H}^{k}-\alpha r\frac{\mathbf{Q}^{k}}{r}||^2_{\text{F}}\\
		=&||\mathbf{\tilde X}^{k+1}-\mathbf{\tilde X}^{k}+\alpha r(\mathbf{\tilde X}^{k}- \mathbf{H}^{k}-\mathcal{C}_r(\mathbf{\tilde X}^{k}- \mathbf{H}^{k}))||^2_{\text{F}}\\
		\leq &\tau_1\Big [\alpha r||\mathbf{\tilde X}^{k}- \mathbf{H}^{k}-\mathcal{C}_r(\mathbf{\tilde X}^{k}- \mathbf{H}^{k})||^2_{\text{F}}\\
		&+(1-\alpha r)||\mathbf{\tilde X}^{k}- \mathbf{H}^{k}||^2_{\text{F}}\Big ]+\frac{\tau_1}{\tau_1-1}||\mathbf{\tilde X}^{k+1}-\mathbf{\tilde X}^{k}||^2_{\text{F}},\\
	\end{split}
\end{equation}
where in the first inequality we use the result of Lemma 5 in \cite{liao2022compressed} with $\tau_1=1+\alpha r \delta$. 

Taking conditional expectation on both sides of \eqref{cpe1}, we obtain
\begin{equation}\label{cpe2}
	\begin{split}
		&\mathbb{E}[||\mathbf{\tilde X}^{k+1}- \mathbf{H}^{k+1}||^2_{\text{F}}\mid \mathcal{F}^k]\\
		&\leq \tau_1[\alpha r (1-\delta)+(1-\alpha r)]\mathbb{E}[||\mathbf{\tilde X}^{k}- \mathbf{H}^{k}||^2_{\text{F}}\mid \mathcal{F}^k]\\
		&+\frac{\tau_1}{\tau_1-1}\mathbb{E}[||\mathbf{\tilde X}^{k+1}-\mathbf{\tilde X}^{k}||^2_{\text{F}}\mid \mathcal{F}^k],\\ 
	\end{split}
\end{equation}
where the inequality holds based on Assumption \ref{c3}. 

Moreover, we have
\begin{equation}\label{cpe3}
	\begin{split}
		&\mathbb{E}[||\mathbf{\tilde X}^{k+1}-\mathbf{\tilde X}^{k}||^2_{\text{F}}\mid \mathcal{F}^k]\\
		&=\mathbb{E}[||\mathbf{X}^{k+1}-\mathbf{X}^{k}-(\bm{\xi}^{k+1}-\bm{\xi}^k)||^2_{\text{F}}\mid \mathcal{F}^k]\\
		&\leq 2\mathbb{E}[||\mathbf{X}^{k+1}-\mathbf{X}^{k}||^2_{\text{F}}\mid \mathcal{F}^k]+2\mathbb{E}[||\bm{\xi}^{k+1}-\bm{\xi}^k||^2_{\text{F}}\mid \mathcal{F}^k]\\
		&\leq 2\mathbb{E}[||\mathbf{X}^{k+1}-\mathbf{X}^{k}||^2_{\text{F}}\mid \mathcal{F}^k]+8n^2\bar \theta^2.\\
	\end{split}
\end{equation}
Next, we bound $\mathbb{E}[||\mathbf{X}^{k+1}-\mathbf{X}^{k}||^2_{\text{F}}\mid \mathcal{F}^k]$.
\begin{equation}\label{cpe4_1}
	\begin{split}
		&\mathbb{E}[||\mathbf{X}^{k+1}-\mathbf{X}^{k}||^2_{\text{F}}\mid \mathcal{F}^k]\\
		&= \mathbb{E}[||\gamma(I-W)\mathbf{E}^{k}-\gamma(\mathbf{F}_a(\mathbf{X}^k)-\mathbf{F}_a(\mathbf{X}^\star))+\bm{\xi}^k||^2_{\text{F}}\mid \mathcal{F}^k]\\
		&\leq 3\gamma^2||(I-W)||^2_{\text{F}} \mathbb{E}[||\mathbf{E}^{k}||^2_{\text{F}}\mid \mathcal{F}^k]\\
		&+3\gamma^2 L_F^2\mathbb{E}[||\mathbf{X}^{k}- \mathbf{X}^{\star}||^2_{\text{F}}\mid \mathcal{F}^k]+3\mathbb{E}[||\bm{\xi}^k||^2_{\text{F}}\mid \mathcal{F}^k] \\
		&\leq 6C\gamma^2||(I-W)||^2_{\text{F}} \mathbb{E}[||\mathbf{\tilde X}^{k}-\mathbf{H}^{k}||^2_{\text{F}}\mid \mathcal{F}^k]\\
		&+3\gamma^2 L_F^2\mathbb{E}[||\mathbf{X}^{k}- \mathbf{X}^{\star}||^2_{\text{F}}\mid \mathcal{F}^k]+ (12\gamma^2||(I-W)||^2_{\text{F}}+6)n^2\bar\theta^2.
	\end{split}
\end{equation}

Bringing \eqref{cpe3} and \eqref{cpe4_1} into \eqref{cpe2} and denoting $c_4=\frac{6(1+\alpha r \delta)}{\alpha r \delta}>1, c_x=\tau_1[\alpha r (1-\delta)+(1-\alpha r)]=1-\alpha^2r^2\delta^2<1$, we have
\begin{equation}\label{cpe4}
	\begin{split}
		&\mathbb{E}[||\mathbf{X}^{k+1}- \mathbf{H}^{k+1}||^2_{\text{F}}\mid \mathcal{F}^k]\\
		&\leq c_4\gamma^2 L^2_F\mathbb{E}[||\mathbf{X}^{k}-\mathbf{X}^{\star}||_{\text{F}}^2\mid \mathcal{F}^k]\\
		&+ (c_x+c_5\gamma^2)\mathbb{E}[||\mathbf{\tilde X}^{k}- \mathbf{H}^{k}||^2_{\text{F}}\mid \mathcal{F}^k]+c_6\bar\theta^2,\\
	\end{split}
\end{equation}
where $c_5=2c_4C||(I-W)||^2_{\text{F}}$ and $ c_6=(2c_4+6)n^2$.

\subsection{Proof of Theorem \ref{th2}}\label{ap2}
From CDP-NES, it is clear that the observation sequence $\mathcal{O} = \{\mathcal{O}^k\}_{k=0}^{K}$ is uniquely determined by the noise sequences $\bm{\xi} = \{\bm{\xi}^k\}_{k=0}^{K}$ and compression randomness sequence $\bm{\zeta} = \{\bm{\zeta}^k\}_{k=0}^{K}$, where $\bm{\zeta}^k =[\bm{\zeta}_1^k,\bm{\zeta}_2^k, \ldots, \bm{\zeta}_n^k]\in \mathbb{R}^{n\times n}$ is a matrix and its row denotes the compression perturbation of $\mathbf{\tilde x}_i^k-\mathbf{h}_i^k$. We use function $\mathcal{Z}_\mathcal{F}$ to denote the relation, i.e., $\mathcal{O} = \mathcal{Z}_\mathcal{F}(\bm{\xi},\bm{\zeta})$, where $\mathcal{F} = \{\mathbf{X}^0, \mathbf{H}^0, W, \mathcal{S}\}$. From Definition \ref{df_dp}, to show the differential privacy of the cost function $J_{i_0}$, we need to show that the following inequality holds for any observation $\mathcal{O}$ and any pair of adjacent cost function sets $\mathcal{S}^{(1)}$ and $\mathcal{S}^{(2)}$, $$
\begin{aligned}
	P\{(\bm{\xi},\bm{\zeta})& \in \Psi \mid \mathcal{Z}_\mathcal{F}^{(1)}((\bm{\xi},\bm{\zeta}) \in \mathcal{O}\}\\
	& \leq e^\epsilon P\{(\bm{\xi},\bm{\zeta}) \in \Psi \mid \mathcal{Z}_\mathcal{F}^{(2)}(\bm{\xi},\bm{\zeta}) \in \mathcal{O}\},
\end{aligned}$$where $\mathcal{F}^{(l)}=\{\mathbf{X}^0, \mathbf{H}^0, W, \mathcal{S}^{(l)}\}$, $l = 1, 2$, and $\Psi$ denotes the sample space.

Then it is indispensable to guarantee $\mathcal{Z}_\mathcal{F}^{(1)}(\bm{\xi},\bm{\zeta}) = \mathcal{Z}_\mathcal{F}^{(2)}(\bm{\xi},\bm{\zeta})$, i.e.,

\begin{equation}\label{eq:prcondition}
	\mathcal{C}(\mathbf{\tilde x}_i^{k,(1)}-\mathbf{h}_i^{k,(1)}, \bm{\zeta}_i^k) = \mathcal{C}(\mathbf{\tilde x}_i^{k,(2)}-\mathbf{h}_i^{k,(2)}, \bm{\zeta}_i^k),
\end{equation}
$\forall i \in V,  0\le k \le K.$

Since $\mathbf{h}^{0,(1)}_i=\mathbf{h}^{0,(2)}_i$, from the dynamics of CDP-NES algorithm, we have $\mathbf{h}^{k,(1)}_i=\mathbf{h}^{k,(2)}_i, 0\leq k\leq K, \forall i \in \mathcal{V}$ if equation \eqref{eq:prcondition} is satisfied.

Then if the following equation holds
\begin{equation}\label{eq:tildex}
	\mathbf{\tilde x}_{i}^{k,(1)} =\mathbf{\tilde x}_{i}^{k,(2)}, \bm{\zeta}_i^{k,(1)}=\bm{\zeta}_i^{k,(2)}, \forall i \in \mathcal{V}, 0\leq k \leq K,
\end{equation}one obtains that
\begin{equation}
	\begin{aligned}
		&f_{c}(\mathbf{\tilde x}_i^{k,(1)}-\mathbf{h}_i^{k,(1)}, \bm{\zeta}_i^{k,(1)}) = f_{c}(\mathbf{\tilde x}_i^{k,(2)}-\mathbf{h}_i^{k,(2)}, \bm{\zeta}_i^{k,(2)}).\\
	\end{aligned}
\end{equation}

	In order to  ensure equation \eqref{eq:tildex}, the Laplace noise should satisfy 
	\begin{equation}\label{xi1}
		\xi_{i}^{k,(1)} = \xi_{i}^{k,(2)}, 0\le k \le K, \forall i \neq i_0.
	\end{equation}
	
	Similarly, for agent $i_0$,  at iteration $k=0$, we have 
	\begin{equation}\label{xi2}
		\xi_{i_0}^{0,(1)} = \xi_{i_0}^{0,(2)}.
	\end{equation}	
	
	For $k\ge1$, since $J_{i_0}^{(1)}\ne J_{i_0}^{(2)}$, the noise should satisfy
	\begin{equation}\label{xi3}
		\Delta \xi_{i_0}^k = -\Delta \mathbf{x}_{(i_0)}^k, 1\leq k \leq K,
	\end{equation}
	where $\Delta \xi_{i_0}^k=\xi_{i_0}^{k,(1)}-\xi_{i_0}^{k,(2)}$ and $\Delta \mathbf{x}_{(i_0)}^k=\mathbf{x}_{(i_0)}^{k,(1)}-\mathbf{x}_{(i_0)}^{k,(2)}$.
	
	In light of dynamics of variable $\mathbf{x}_{(i)}^k$ in CDP-NES and Assumption \ref{bounded_grad}, we have for all $k\geq 1$
	\begin{equation}\label{ine:deltax}\begin{aligned}
			\|\Delta \mathbf{x}_{(i_0)}^k\|_1&\leq \|\gamma \eta\nabla_{i_0} J_{i_0}(\mathbf{x}_{(i_0)}^{k-1,(1)})-\gamma \eta\nabla_{i_0} J_{i_0}(\mathbf{x}_{(i_0)}^{k-1,(2)})\|_1 \\
			&\leq 2\gamma \eta M.
		\end{aligned}
	\end{equation}
	Denote $\mathcal{R}^{(l)}=\{(\bm{\xi}^{(l)},\bm{\zeta}^{(l)})|\mathcal{Z}_{\mathcal{F}^{(l)}}(\bm{\xi}^{(l)},\bm{\zeta}^{(l)}) \in \mathcal{O}\}$, $l=1,2,$ then
	\begin{equation}
		\begin{aligned}
			&P\{(\bm{\xi}^{(l)},\bm{\zeta}^{(l)})\in\Psi | \mathcal{Z}_{\mathcal{F}^{(l)}}(\bm{\xi}^{(l)},\bm{\zeta}^{(l)}) \in \mathcal{O}\}\\
			&=\int_{\mathcal{R}^{(l)}} f_{\xi}(\bm{\xi}^{(l)})f_{c}(\mathbf{\tilde X}^{(l)}-\mathbf{H}^{(l)}, \bm{\zeta}^{(l)})d\bm{\xi}^{(l)}d\bm{\zeta}^{(l)},
		\end{aligned}
	\end{equation}
	where $f_{\xi}(\bm{\xi}^{(l)})=\prod_{k=0}^K\prod_{i=1}^n\prod_{r=1}^n f_{L}([\xi_{i}^{k,(l)}]_r,\theta_i)$ and $f_{c}(\mathbf{\tilde X}^{(l)}-\mathbf{H}^{(l)}, \bm{\zeta}^{(l)})=\prod_{k=0}^K\prod_{i=1}^n f_{c}(\mathbf{\tilde x}_i^{k,(l)}-\mathbf{h}_i^{k,(1)}, \bm{\zeta}_i^{k,(l)})$
	
	According to the above relation \eqref{xi1}-\eqref{xi3}, we can obtain for any $\bm{\xi}^{(1)}$, there exists a  $\bm{\xi}^{(2)}$ such that $\mathcal{Z}_{\mathcal{F}^{(1)}}(\bm{\xi}^{(1)},\bm{\zeta}^{(1)})=\mathcal{Z}_{\mathcal{F}^{(2)}}(\bm{\xi}^{(2)},\bm{\zeta}^{(2)})$. As the converse argument is also true, the above defines a bijection.  Hence, for any $\bm{\xi}^{(2)}$, there exists a unique $(\bm{\xi}^{(1)}, \Delta\bm{\xi})$	such that $\bm{\xi}^{(2)}=\bm{\xi}^{(1)}+\Delta\bm{\xi}$. Since $\Delta\bm{\xi}$ is fixed and is not dependent on $\bm{\xi}^{(2)}$, we can use a change of variables to obtain 
	$		P\{(\bm{\xi}^{(2)},\bm{\zeta}^{(2)})\}\in \mathcal{R}^{(2)}\}	=\int_{\mathcal{R}^{(1)}} f_{\xi}(\bm{\xi}^{(1)}+\Delta \bm{\xi})f_{c}(\mathbf{\tilde X}^{(1)}-\mathbf{H}^{(1)}, \bm{\zeta}^{(1)})d\bm{\xi}^{(1)}d\bm{\zeta}^{(1)}.
	$
	
	Hence, we have
	\begin{equation}
		\begin{aligned}
			&\frac{P\{(\bm{\xi},\bm{\zeta})\in\Psi |\mathcal{Z}_{\mathcal{F}^{(1)}}(\bm{\xi},\bm{\zeta}) \in \mathcal{O}\}}{P\{(\bm{\xi},\bm{\zeta})\in\Psi |\mathcal{Z}_{\mathcal{F}^{(2)}}(\bm{\xi},\bm{\zeta}) \in \mathcal{O}\}}=\frac{P\{(\bm{\xi}^{(1)},\bm{\zeta}^{(1)})\in \mathcal{R}^{(1)}\}}{P\{(\bm{\xi}^{(2)},\bm{\zeta}^{(2)})\in \mathcal{R}^{(2)}\}}\\
			&=\frac{\int_{\mathcal{R}^{(1)}} f_{\xi}(\bm{\xi}^{(1)})f_{c}(\mathbf{\tilde X}^{(1)}-\mathbf{H}^{(1)}, \bm{\zeta}^{(1)})d\bm{\xi}^{(1)}d\bm{\zeta}^{(1)}}{\int_{\mathcal{R}^{(1)}} f_{\xi}(\bm{\xi}^{(1)}+\Delta \bm{\xi})f_{c}(\mathbf{\tilde X}^{(1)}-\mathbf{H}^{(1)}, \bm{\zeta}^{(1)})d\bm{\xi}^{(1)}d\bm{\zeta}^{(1)}}.\\
		\end{aligned}
	\end{equation}
	Since 
	\begin{equation}
		\begin{aligned}
			&\frac{ f_{\xi}(\bm{\xi}^{(1)})d\bm{\xi}^{(1)}}{\ f_{\xi}(\bm{\xi}^{(1)}+\Delta \bm{\xi})d\bm{\xi}^{(1)}}
			=\prod_{k=0}^K\prod_{i=1}^n\prod_{r=1}^n \frac{f_{L}([\xi_{i}^{k,(1)}]_r,\theta_i)}{f_{L}([\xi_{i}^{k,(1)}+\Delta \xi_{i}^k]_r,\theta_i)}\\
			&\leq \prod\limits_{k=0}^K\prod\limits_{r=1}^n\exp\Big(
			\frac{\Big|[\Delta \xi_{i_0}^k]_r\Big|}{\theta_{i_0}}\Big)\\
			&=\exp\Big(\sum\limits_{k=0}^K\frac{||\Delta \xi_{i_0}^k||_1}{\theta_{i_0}}\Big)\leq\exp\Big(\frac{2\gamma\eta KM}{\theta_{i_0}}\Big)= e^{\epsilon_{i_0}},
		\end{aligned}
	\end{equation}
	we have 
	\begin{equation}
		\begin{aligned}
			P\{(\bm{\xi},\bm{\zeta})\in&\Psi |\mathcal{Z}_{\mathcal{F}^{(1)}}(\bm{\xi},\bm{\zeta}) \in \mathcal{O}\}\\
			&\le e^{\epsilon_{i_0}} P\{(\bm{\xi},\bm{\zeta})\in\Psi |\mathcal{Z}_{\mathcal{F}^{(2)}}(\bm{\xi},\bm{\zeta}) \in \mathcal{O}\},\\
		\end{aligned}
	\end{equation}	 
	which establishes the $\epsilon_{i_0}$-differential privacy  of agent $i_{0}$. The fact that $i_0$ can be  arbitrary  without loss of generality, guarantees $\epsilon_i$-differential privacy of each agent $i$.

\end{document}